\documentclass[prl,twocolumn,citeautoscript,amsmath]{revtex4-1}

\usepackage{graphicx}
\usepackage{newtxtext,newtxmath}
\usepackage{epsfig}
\usepackage{amsmath,bbm,amssymb,bm,times}
\usepackage[colorlinks,linkcolor=blue,citecolor=blue]{hyperref}

\newcommand{\eqn}[1]{
\begin{eqnarray}
	#1
\end{eqnarray}
}

\begin{document}

\title{Learning the best thermoelectric nanoscale heat engines 
through evolving network topology}

\author{Yuto Ashida}
\email{ashida@ap.t.u-tokyo.ac.jp}
\affiliation{Department of Applied Physics, University of Tokyo, 7-3-1 Hongo, Bunkyo-ku, Tokyo 113-8656, Japan}
\author{Takahiro Sagawa}
\affiliation{Department of Applied Physics, University of Tokyo, 7-3-1 Hongo, Bunkyo-ku, Tokyo 113-8656, Japan}

\begin{abstract}
The quest to identify the best heat engine has been at the center of science and technology. Thermoelectric nanoscale heat engines convert heat flows into useful work in the form of electrical power and promise the realization of on-chip power production. Considerable studies have so far revealed the potentials to yield an enhanced efficiency originating from quantum confinement effects and energy-dependent transport properties. However, the full benefit of many-body interactions in thermoelectric is yet to be investigated; identifying the optimal interaction is a hard problem due to combinatorial explosion of the search space, which makes brute-force searches infeasible. Here we tackle this problem with reinforcement learning of network topology in interacting electronic systems, and identify a set of the best thermoelectric nanoscale engines. Harnessing many-body interactions, we show that the maximum possible values of the thermoelectric figure of merit and the power factor can be enhanced by orders of magnitudes for generic single-electron levels. This allows for simple and flexible design of realizing the asymptotic Carnot efficiency with subextensive, but still nonzero and stable power. To realize the optimal nanoscale engines, we propose concrete physical setups based on quantum-dot arrays. The developed framework of reinforcement learning through evolving network topology thus enables one to identify full potential of nanoscale systems.
\end{abstract}
\maketitle
\begin{figure*}
\includegraphics[width=135mm]{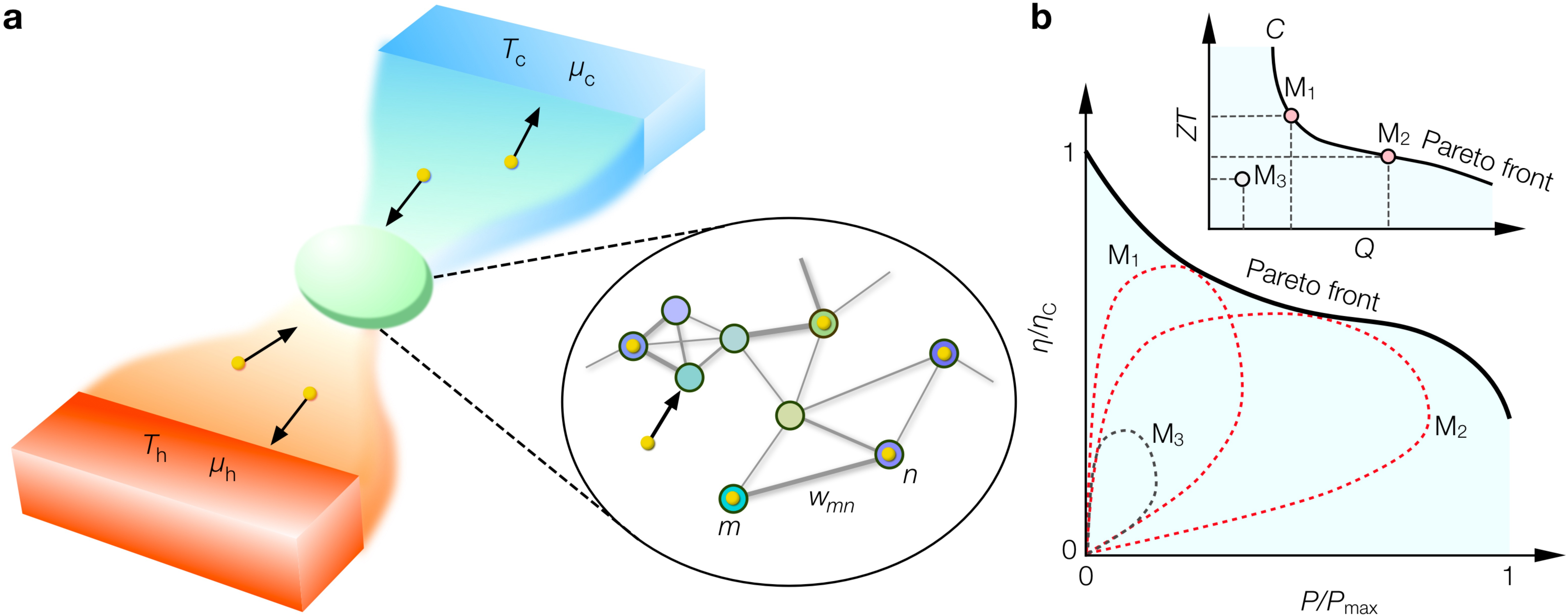} 
\caption{\label{fig1}
Graphical representation of interacting nanothermoelectric. (a) Thermoelectric nanoscale heat engine is characterized by a network, in which each node represents a single-particle level labeled by an integer like $m,n$. Each edge connecting two nodes indicates the presence of interaction between the corresponding single-particle levels (in, e.g., quantum dots). The width of the edge between node $m$ and node $n$ is labeled by $w_{mn}$ and represents the interaction strength. The system is connected with hot (h) and cold (c) reservoirs at temperatures $T_{\rm h,c}$ and electrochemical potentials $\mu_{\rm h,c}$. Each node exchanges electrons with the reservoirs and can be occupied by at most a single electron.  (b) Set of the best heat engines is characterized by the best tradeoff (known as the Pareto front) between their power $P$ and efficiency $\eta$ (thick solid curve in the main panel). The inset illustrates the concept of dominance in terms of  power factor $Q$ and  figure of merit $ZT$; machines ${\rm M}_{1,2}$ dominate ${\rm M}_3$ while there are no dominance relations between ${\rm M}_{1}$ and ${\rm M}_2$. The Pareto front $\cal C$ in the inset is defined by the Pareto-optimal machines (such as ${\rm M}_{1}$ and ${\rm M}_2$) that are not dominated by any other ones.
The dashed loops in the main panel indicate the power-efficiency curves with varying chemical potentials at fixed $Q$ and $ZT$ (cf. Eq.~\eqref{tradeoff}), which correspond to machines ${\rm M}_{1,2,3}$ indicated in the inset. The envelope of all the possible loops provides the Pareto front on the $P$-$\eta$ plane in the main panel.
}
\end{figure*}

In 1824, Sadi Carnot found \cite{SC72} that the efficiency of heat engines operating between a hot reservoir at temperature $T_{\rm h}$ and a cold one at $T_{\rm c}$ is universally bounded from above by the value
\eqn{
\eta_{{\rm C}}=1-\frac{T_{{\rm c}}}{T_{{\rm h}}},
}
which is now known as the Carnot efficiency. This limit can be reached for ideal reversible machines operating at quasistatic conditions, leading to infinite operation-time and vanishing output power. In contrast, any useful devices must supply nonvanishing power that necessarily associates with nonzero entropy productions and  efficiency below $\eta_{\rm C}$. Thus, it is of both fundamental and practical importance to consider what is the best heat engine with finite power. 
In this direction, a seminal work was done by Curzon and Ahlborn, who have considered \cite{FC75} a thermal machine operating at the maximum power as the optimal engine. In practice, a promising candidate for realizing energy-efficient power production is nanostructured thermoelectric \cite{CJV10,YD11,MM13,BS15}, which converts heat flows into electrical power (see Fig.~\ref{fig1}a). It has attracted considerable interest as promoted by the prospect of an enhanced efficiency due to quantum confinement effects \cite{LDH93,TCH02} and significant reduction of the phonon thermal conductivity \cite{RV01,AIB08}.  

Since efficiency and power are two conflicting objectives, one cannot in general find a heat engine that optimizes both of them simultaneously. This class of problems is known as the multiobjective optimization problem for which the solution is given by the quantitative identification of tradeoff between multiple objectives \cite{YS85}. 
More specifically, in this paper we propose to characterize the best heat engines by a set of machines for which efficiency and power cannot be further improved without compromising the other (see the main panel in Fig.~\ref{fig1}b). In the language of the multiobjective optimization theory, such type of set is known as the {\it Pareto front} \cite{YS85}. From this viewpoint, the engines considered by Carnot \cite{SC72} and Curzon-Ahlborn \cite{FC75} are two specific examples of a more general set of the multiobjective-optimal heat engines.

Thermoelectric heat engines in the linear-response regime can be fully characterized by the figure of merit $ZT$ and the power factor $Q$, which are related to the maximum possible values of efficiency and power, respectively. Thus, we can reduce a problem of seeking the best thermoelectric to the search of the Pareto front on the objective space spanned by $ZT$ and $Q$ (see the inset in Fig.~\ref{fig1}b). 
In noninteracting systems, brute-force optimizations of transport functions have allowed one to identify the best thermoelectric achieving the highest $ZT$ \cite{MGD96} and, more generally, the optimal power-efficiency tradeoff \cite{RSW15}. 
However, the challenging goal of identifying the best {\it interacting} thermoelectric is still unexplored.

Currently, a major challenge in nanoscale thermoelectric is that an individual thermal machine can only supply low power output. To realize the promise of on-chip power production, it is indispensable to assemble a large number of nanoscale engines, where the Coulomb interaction among the constituents becomes inevitable due to its long-range nature.   
There is thus a strong need to reveal full potential of many-body interaction on thermoelectric performances. Does the interaction enhance efficiency and power and, if yes, to what extent? Can one identify the best possible interacting thermal machines? 
Here we answer these questions in the affirmative by developing a framework for reinforcement learning of network topology in interacting electronic systems. The key ideas are to map the many-body interaction among electrons onto the network (see Fig.~\ref{fig1}a), and to train its topology through the differential evolution, which is one of the most competitive training algorithms in high-dimensional nonconvex search space \cite{RS95,FN09}. 

{\it Interacting systems linked with network topology.---} 
We consider an arbitrary fermionic system described by the many-body Hamiltonian
\eqn{\label{Ham}
{H}=\sum_{l}(\epsilon_{l}-v_{g})n_{l}+\frac{1}{2}\sum_{l\neq m}w_{lm}n_{l}n_{m},
}
where $\epsilon_l$ denotes single-particle energy of mode $l=1,2,\ldots,N_f$, $v_g$ is the ground voltage, $n_l=0,1$ represents the occupation of zero or one electron, and $w_{lm}\geq 0$ are generic two-body interaction parameters. 
The system is in contact with hot (h) and cold (c) reservoirs at temperatures $T_{\rm h,c}$ and electrochemical potentials $\mu_{\rm h,c}$ (see Fig.~\ref{fig1}a). We consider the sequential regime, in which transport occurs due to single-electron tunnelings to reservoirs and generation of quantum coherence can be neglected. Then, the dynamics can be described by the classical master equation \cite{CWB90}. The tunneling rates to reservoirs are assumed to be energy independent and denoted by $\gamma_{\rm h,c}>0$.
Denoting the particle current and the heat flow out of each reservoir as $J^i$ and $J^i_q$ with $i={\rm h,c}$, the power and the efficiency are given by $P=-\sum_{i}\mu_i J^i$ and $\eta=P/J_q^{\rm h}$, respectively.

Our aim is to establish which of electronic systems modeled by Eq.~\eqref{Ham} can fundamentally achieve the best power-efficiency tradeoff. To this end, we take into account all the possible (repulsive) interactions $w$ while the phonon contribution to heat flow is not included as it is external to the electronic system \cite{CJV10}.
The present model can thus be graphically represented as a network as shown in Fig.~\ref{fig1}a. Here, each node indicates a single-particle level (in, e.g., quantum dots) that exchanges  electrons with reservoirs. An edge between two nodes represents the presence of interaction between the corresponding single-particle levels, and its width indicates the strength of the interaction. From this perspective, the problem of identifying the best heat engines via optimizing (i.e., training) parameters $w$ can be considered as reinforcement learning of underlying topology and weight values of the interaction network.

\begin{figure*}
\includegraphics[width=140mm]{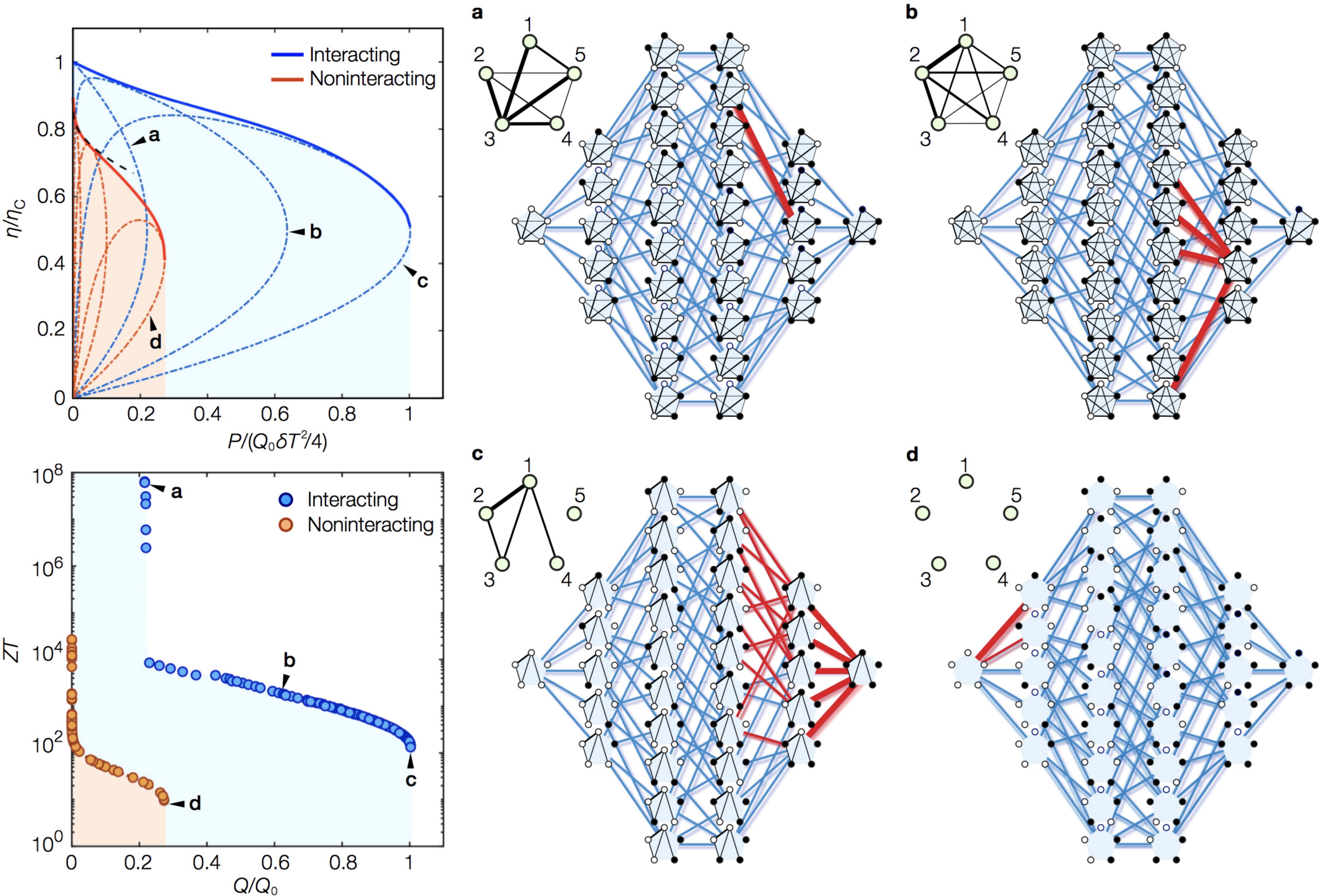} 
\caption{\label{fig2}
Learning network topologies of the best nanoscale heat engines. 
The left top (bottom) panel shows the obtained best tradeoff between power $P$ and  efficiency $\eta$ (power factor $Q$ and figure of merit $ZT$).
The blue (red) shaded region is allowed for interacting  (noninteracting) heat engines and its upper boundary represents the Pareto front, i.e., the set of the optimal machines for which two objectives cannot be further improved without compromising the other. 
The black dashed curve in the left top panel indicates the scaling $1-\eta/\eta_{\rm C}\propto\sqrt{P}$ at low power in the noninteracting case. 
 (a)-(d)  Topologies of the interaction networks (insets) and the state-transfer networks (main panels) for the Pareto-optimal heat engines indicated by labels  (a)-(d) in the left panels. They attain the highest $ZT$ (a), the suboptimal $ZT$ and $Q$ (b), the highest $Q$ (c), and the highest $Q$ in the noninteracting case (d). The red edges in the state-transfer networks indicate the activated transfer edges along which the probability flow is significant. 
 In (c), the highest power is achieved by the sparse interaction network, where  degenerate weakly interacting holes result in the maximal activation of transfer edges. 
In (a) and (b), the strong and dense interactions isolate several states from the other ones and solely activate transfer edges within those isolated manifolds. This enhances $ZT$ at the expense of compromising $Q$ (see the left bottom panel).
The divergence of $ZT$ at $Q/Q_0\simeq 0.21$ originates from the almost perfect unicyclic structure (see panel (a)), leading to the tight-coupling condition. 
We set $N_f=5$, $\epsilon_{l}=l\Delta$ with $\Delta/k_{\rm B}T=3$, $\gamma_{\rm h}=\gamma_{\rm c}\equiv\gamma$ and plot $Q$ in the unit of $Q_{0}=k_{\rm B}\gamma/T$.
}
\end{figure*}

{\it Power-efficiency tradeoff and Pareto-optimal thermal machines.---} 
To be specific, hereafter we focus on the linear-response regime, $\delta T=T_{\rm h}-T_{\rm c}\ll T_{\rm h}$ and $|\delta\mu|=|\mu_{\rm h}-\mu_{\rm c}|\ll k_{\rm B}T_{\rm h}$, and denote $T=T_{\rm h}\simeq T_{\rm c}$. 
The thermoelectric properties are then characterized by the figure of merit $ZT$ and the power factor $Q$ that are defined by
\eqn{
ZT=\frac{\sigma S^{2}T}{\kappa}=\frac{QT}{\kappa}.
} 
Here, $\sigma$ is the electrical conductance, $S$ is the Seebeck coefficient, and $\kappa$ is the thermal conductance; their values can be associated with the Onsager coefficients (see Supplementary Materials). More explicitly, the power-efficiency tradeoff can be fully characterized by $ZT$ and $Q$ via the linear-response formula \cite{GB17}
\eqn{\label{tradeoff}
\frac{\eta(P)}{\eta_{{\rm C}}}=\frac{P/(Q\delta T^{2}/4)}{2\left[1+2/ZT\mp\sqrt{1-P/(Q\delta T^{2}/4)}\right]},
}
which have two branches as they correspond to changing $\delta \mu$ from zero to  $S\delta T$ at which $P=0$ again; the stopping value $S\delta T$ corresponds to the point at which the sign of the electron flow reverses. It follows from Eq.~\eqref{tradeoff} that $ZT$ and $Q$ characterize the maximum possible values of the efficiency $\eta/\eta_{\rm C}\leq(\sqrt{ZT+1}-1)/(\sqrt{ZT+1}+1)$ and the power $P\leq Q\delta T^2/4$, respectively. 

The problem of finding the best heat engines now reduces to the multiobjective optimization of $ZT$ and $Q$. To this end, let ${\rm M}_{\cal W}$ symbolically represent a thermoelectric machine characterized by a set of $1\!+\!N_f(N_f\!-\!1)/2$ variables ${\cal W}=\{v_g,w\}$, which includes the ground voltage and interaction parameters while the single-particle energies are assumed to be given. We call that a machine ${\rm M}_{\cal W}$ {\it dominates} \cite{YS85} ${\rm M}_{{\cal W}'}$ (denoted as ${\rm M}_{\cal W}\succ {\rm M}_{{\cal W}'}$) if ${\rm M}_{\cal W}$ is no worse than ${\rm M}_{{\cal W}'}$ in both $ZT$ and $Q$, and ${\rm M}_{\cal W}$ is strictly better than ${\rm M}_{{\cal W}'}$ in at least one of them. A machine ${\rm M}_{\cal W^{*}}$ is {\it Pareto-optimal} if no other machines dominate it. Then, the Pareto front \cite{YS85} $\cal C$ is the curve defined by a set of $(Q({\cal W^{*}}),ZT({\cal W^{*}}))$ for all the possible Pareto-optimal machines. 
We illustrate these concepts in the inset of Fig.~\ref{fig1}b, where the machines satisfy ${\rm M}_{1,2}\succ {\rm M}_{3}$ while there are no dominance relations between ${\rm M}_{1}$ and ${\rm M}_{2}$.

We emphasize that optimizing $ZT$ alone is insufficient as $ZT$ has no information about the maximum possible power, which is crucial for realizing any useful devices. We instead characterize the best thermoelectric in terms of the Pareto front $\cal C$ on the $Q$-$ZT$ plane, which allows us to identify the full set of the optimal engines at finite power. The corresponding Pareto front on the power-efficiency plane can be also given as the envelope of the image space of ${\cal C}$ through the mapping~\eqref{tradeoff} (cf. the main plane in Fig.~\ref{fig1}b). 

We are now left with the task of finding the best nanoscale engines by using machine learning to train the network parameters ${\cal W}$.
However, even after the above simplifications, the problem still remains challenging as the transport coefficients $\sigma$, $S$, and $\kappa$ are adversely interdependent while brute-force approaches become quickly infeasible. The latter is  due to the exponential growth $\sim (L_{\rm dis})^{N_f^2/2}$ of the search space with the system size $N_f$, where $L_{\rm dis}$ is the number of bins for discretizing a continuous parameter. Moreover, the greedy (gradient-based) algorithms overwhelmingly fail even for few-level systems because of the proliferation of local optima (see Supplementary Materials). 
To overcome the challenges, we employ a global search approach based on the differential evolution \cite{RS95}, which is one of the most powerful gradient-free algorithms inspired by a process whereby biological organisms adapt and survive. The key advantage is its autonomous adaption of the balance between exploitation and exploration, which can significantly expedite an efficient search over a high-dimensional nonconvex landscape \cite{FN09} (see Supplementary Materials). 

\begin{figure*}
\includegraphics[width=140mm]{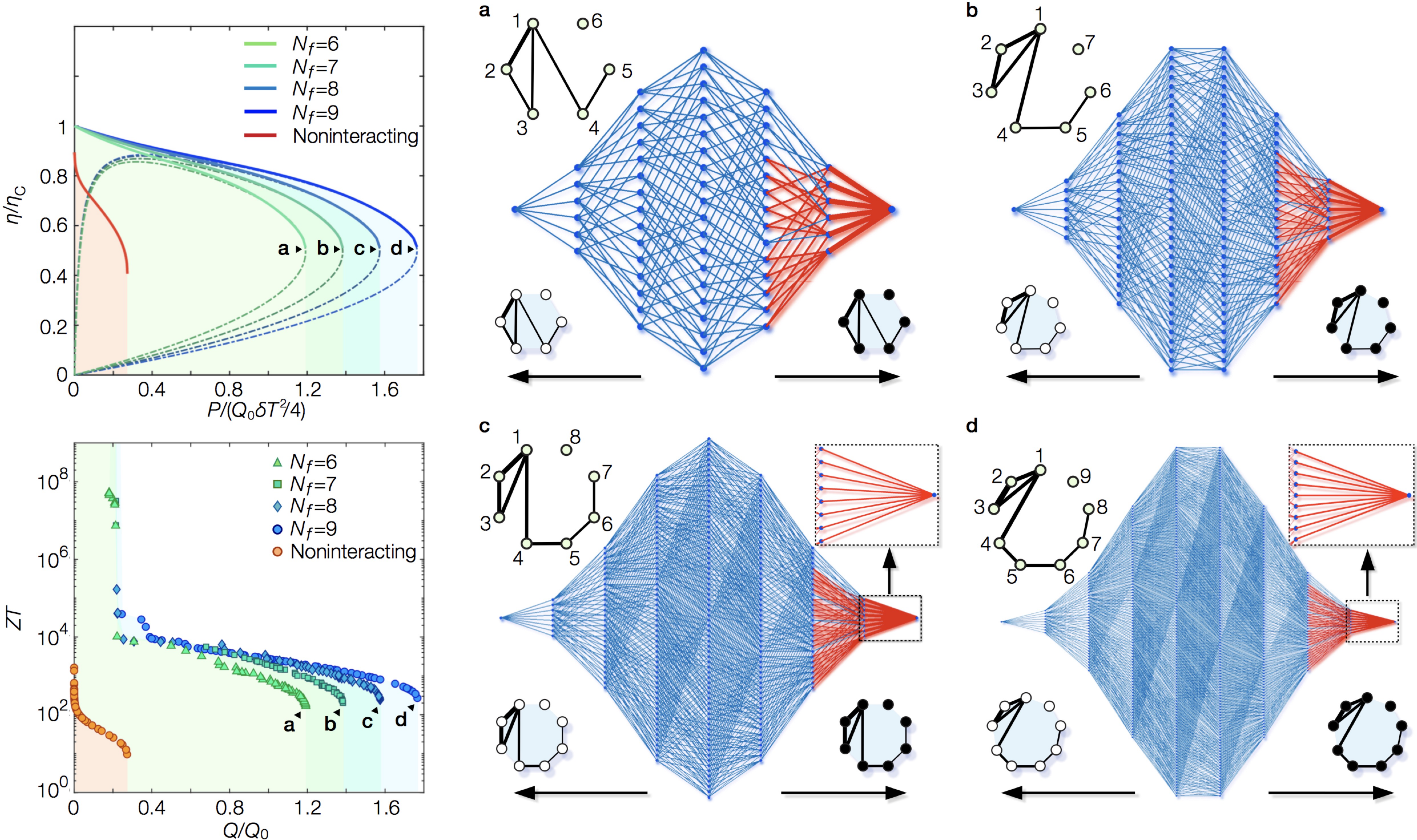}
\caption{\label{fig3}
Size dependence of the highest-power nanoscale heat engine. 
The left top panel shows the best tradeoffs between power $P$ and efficiency $\eta$ for the highest-power heat engines at different system sizes $N_f$. The left bottom panel shows the corresponding Pareto fronts in terms of power factor $Q$ and figure of merit $ZT$. 
In both panels, the allowed regions (color shaded) expand with increasing $N_f$.  
(a)-(d) Topologies of the interaction networks (insets) and the state-transfer networks (main panels) for the highest-power heat engines with $2^{N_f}=64$ (a), $128$  (b), $256$ (c), and $512$ states (d). 
The red edges in the state-transfer networks indicate the activated transfer edges associated with significant probability flows. 
All of these highest-power engines share the common features; the interaction network is optimized such that single-hole excitation energies are degenerate and its topology is sparse as much as possible. The former leads to the activation of all the transfer edges  between the fully occupied state and single-hole states (see also magnified insets in  (c,d)). The latter eliminates unfavorable interactions among hole excitations, leading to the maximal activation of transfer edges between single-hole states and double-hole states. For the sake of comparison, the data of the noninteracting case for $N_f=9$ are given in the left panels while the results are almost independent of $N_f$. The other parameters are set to be the same as in Fig.~\ref{fig2}.
}
\end{figure*}

{\it Learning the power-efficiency tradeoff in nanoscale thermoelectric.---} 
The left bottom panel in Fig.~\ref{fig2} shows the Pareto front on the $Q$-$ZT$ plane, which is identified by the above learning protocol. With many-body interaction and generic (nondegenerate) single-electron levels, the values of $Q$ and $ZT$ can be enhanced by orders of magnitudes in comparison with those in the noninteracting case. As a consequence, interaction can significantly improve power-efficiency tradeoff in heat-to-work conversion (see the left top panel in Fig.~\ref{fig2}). 
The substantial enhancements originate from the ability of interaction to activate many paths of state transitions as well as attain the approximate tight-coupling condition $J\propto J_{q}$. The former leads to a much higher $Q$ than that in the noninteracting case while the latter can feature a significantly large $ZT$ \cite{JS76,US12}.

To further elucidate this physical mechanism, we visualize typical realizations of the Pareto-optimal heat engines in Fig.~\ref{fig2}a-d. Here, the pentagon-shaped, inset networks show the identified optimal interactions, where 
nodes (edges) represent $N_f$ single-electron levels (interactions between them). The networks in the main panels of Fig.~\ref{fig2}a-d visualize the whole state-transfer networks with nodes (edges) being $2^{N_f}$ many-body states (transitions between them). 

The inset network in Fig.~\ref{fig2}c demonstrates that the highest power has been achieved by a surprisingly sparse interaction. 
As shown in the state-transfer network (the main panel of Fig.~\ref{fig2}c), probability flows concentrate on the transfer edges among the fully occupied state, single- and two-hole states as highlighted by red color. 
This activation of many transfer paths originates from the degeneracies of single-hole states and the suppression of unfavorable hole-hole interactions, both of which are achieved by the observed sparse interaction. In this way, the power is maximized by activating transfer edges as many as possible.

As making the interaction network stronger and denser, the Pareto-optimal machines can improve $ZT$ (i.e., efficiency) at the expense of compromising $Q$ (i.e., power). This can be inferred from the state-transfer networks in Fig.~\ref{fig2}a,b, where the strong and dense repulsive interaction isolates a particular energy manifold from the other many-body states, realizing the approximate tight-coupling condition $J\propto J_{q}$ within that manifold. This emergent unicyclic structure in the probability flow  allows for high $ZT$; yet, it comes at the price of sacrificing $Q$ due to the reduced number of activated transitions. 

The divergence of $ZT$ at $Q/Q_0\simeq 0.21$  (cf. the left bottom panel in Fig.~\ref{fig2}) originates from an almost perfect unicyclic structure \cite{JS76,US12}. Here, the dense interaction isolates two many-body states from the other ones such that only the probability flow between these two levels is significant (cf. the main panel in Fig~\ref{fig2}a). This ensures the tight-coupling condition with great accuracy and thus leads to the divergent  $ZT$. 
We remark that the all-to-all mean-field coupling (i.e., $w_{lm}=w_{\rm MF}$ for all $l\neq m$), which has been mainly discussed in the previous literature \cite{CWB90,XZ07,JL10,PT12,AI15,PAE17,HV17,TH18}, does not give the Pareto-optimal solutions here. In general, we find that this type of uniform, maximally dense interaction is largely suboptimal. 

Figure~\ref{fig2}d demonstrates that the noninteracting machine with nondegenerate single-electron levels activates only a few edges, resulting in low power output due to the bipolar effect, i.e., a nonzero heat conduction at the zero particle current. 
 In the left top panel in Fig.~\ref{fig2}, it is worthwhile to mention that the obtained Pareto front in the noninteracting (resp. interacting) case is consistent with the scaling at low power, $1-\eta/\eta_{\rm C}\propto P^a$ with $a=1/2$ (resp. $a=1$), which has been previously discussed based on the Landauer-B{\"u}ttiker theory \cite{RSW15} and the molecular simulations \cite{GB14,RL18}.  

Based on these observations, we now conjecture that the Pareto-optimal thermal machine at the highest power can be generally achieved by satisfying the following conditions: (i) single-hole excitation energies $e_l=\epsilon_l+\sum_{m\neq l}w_{lm}$ are degenerate, i.e., $|e_l-e_{l+1}|\ll k_{\rm B}T$ for $l=1,2,\ldots,N_f-1$, (ii) at most $N_f-1$ variables of $\{w_{lm}\}_{l>m}$ can be nonzero, and (iii) the ground voltage is set to be $v_{g}=e_{\rm h}+\alpha k_{\rm B}T$, where $e_{\rm h}$ is the degenerate single-hole energy and $\alpha\simeq 2.40$. 
Physically, the first condition ensures that all the transfer edges between the fully occupied state and single-hole states are activated. The second one makes interactions among hole excitations as sparse as possible, leading to the maximal activation of transfer edges between single-hole states and double-hole states. The final one optimizes the ground voltage in such a way that the power is maximized \cite{PM08,ME09,CVB05}. We confirm the conjecture up to a system with $2^{N_f}=512$ states as demonstrated in Fig.~\ref{fig3}a-d, where the concrete examples of the highest-power machines are shown. 
We note that, for any set $\{\epsilon_l\}$, there exist an excessive number of solutions for interaction parameters satisfying the above conditions. This fact significantly enhances design flexibility of realizing the highest-power heat engines. Meanwhile, in the noninteracting machine, the highest power can be attained only if  single-electron levels are perfectly degenerate \cite{MGD96}, $\epsilon_1\!=\!\epsilon_2\!=\!\cdots\!=\!\epsilon_{N_f}$, which might be challenging to realize due to, e.g., inherent size fluctuations of quantum dots.

\begin{figure}[t]
\includegraphics[width=55mm]{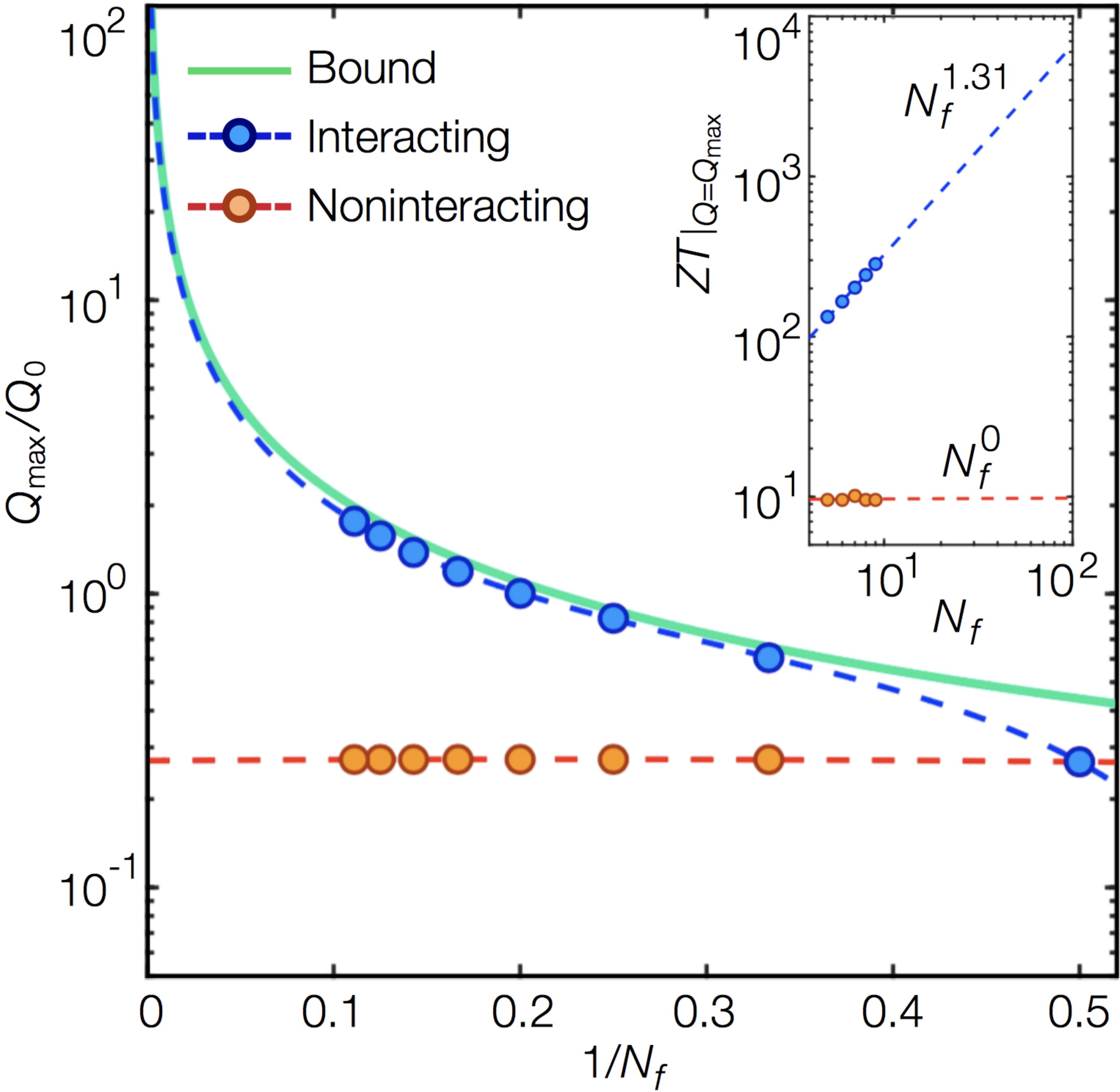}
\caption{\label{fig4}
Finite-size scaling of the maximum possible power.  The blue dashed curve shows an extrapolation of the highest power factors at different sizes $N_f$ using the form $Q_{\rm max}/Q_{0}=\sum_{i=1}^{4}a_i N_f^{i-3}$ with fitting parameters $a_{1,2,3,4}$. It asymptotically achieves the fundamental bound (cf. Eq.~\eqref{bound}) indicated by the green solid curve. The inset plots the associated figure of merit $ZT|_{Q=Q_{\rm max}}$ of the highest-power machine against $N_f$, where the blue dashed line shows the fitted scaling $ZT|_{Q=Q_{\rm max}}\propto N_f^{1.31}$. For the sake of comparison, the noninteracting results are shown in both plots with the red dashed lines being the fitted $N_f$-independent extrapolations. The parameters are set to be the same as in Fig.~\ref{fig2}.
}
\end{figure}

The left panels in Fig.~\ref{fig3} demonstrate that the maximum possible values of efficiency and power increase with the number of levels $N_f$. To further investigate the size dependence, in Fig.~\ref{fig4} we perform the finite-size scaling of the maximum possible power factor $Q_{\rm max}$ and the associated figure of merit $ZT|_{Q=Q_{\rm max}}$. It is clear that both quantities grow with system size $N_f$. In particular, in the thermodynamic limit $Q_{\rm max}$ diverges in proportion to $N_f$, and the power per level asymptotically attains the fundamental bound set by an ideal unicyclic system \cite{ME09}
\eqn{\label{bound}
\frac{Q_{\rm max}}{N_f}\to \xi \frac{k_{\rm B}}{T}\frac{\gamma_{\rm h}\gamma_{\rm c}}{\gamma_{\rm h}+\gamma_{\rm c}}\;\;\;\;(N_f\to \infty),
}
where $\xi\simeq 0.439$. For the noninteracting case with generic (nondegenerate) single-electron levels, both $Q$ and $ZT$ do not grow with $N_f$ and remain to be small finite values even if the number of available energy levels diverges. This result is qualitatively consistent with the previous finding based on the Landauer-B{\"u}ttiker theory \cite{RSW14}.

{\it The asymptotic Carnot efficiency at nonzero power.---}
 We note that the diverging power factor $Q_{\rm max}\propto N_f$ discussed above suggests a simple pathway to asymptotically realize the Carnot efficiency at nonzero power. It has been argued that such a heat engine accompanies divergent fluctuation of power in steady-state regimes \cite{TRG16,PP18} or under cyclic protocols at the criticality \cite{MC16,VH17}. Apparently, these observations might 
suggest that the Carnot efficiency and the stable power (i.e., without too much fluctuations) are incompatible. Yet, we point out that the maximum-power machine above can be still used as a faithful engine as long as the mean power satisfies $P\propto N_f^\zeta$ with $1/2<\zeta<1$ because its fluctuation scales with a weaker exponent as $\delta P\propto\sqrt{\sigma}\propto\sqrt{N_f}$. As inferred from Eq.~\eqref{tradeoff}, the diverging $Q$ and $ZT$ of the highest-power engines then enable the asymptotic Carnot efficiency $\eta(P)\to\eta_{\rm C}$ in the limit of $N_f\to\infty$ with sustaining subextensive, but still stable power $P\gg\delta P$. The price one must pay is precise control of the chemical potentials according to $1-\delta\mu/(S\delta T)\propto P/Q\propto 1/N_f^{1-\zeta}$. We remark that our consideration does not contradict with a known tradeoff \cite{NS16,MP17}, such as $P\leq M(1-\eta/\eta_{\rm C})$, since a constant $M$ is diverging in the present case.

\begin{figure*}[t]
\includegraphics[width=125mm]{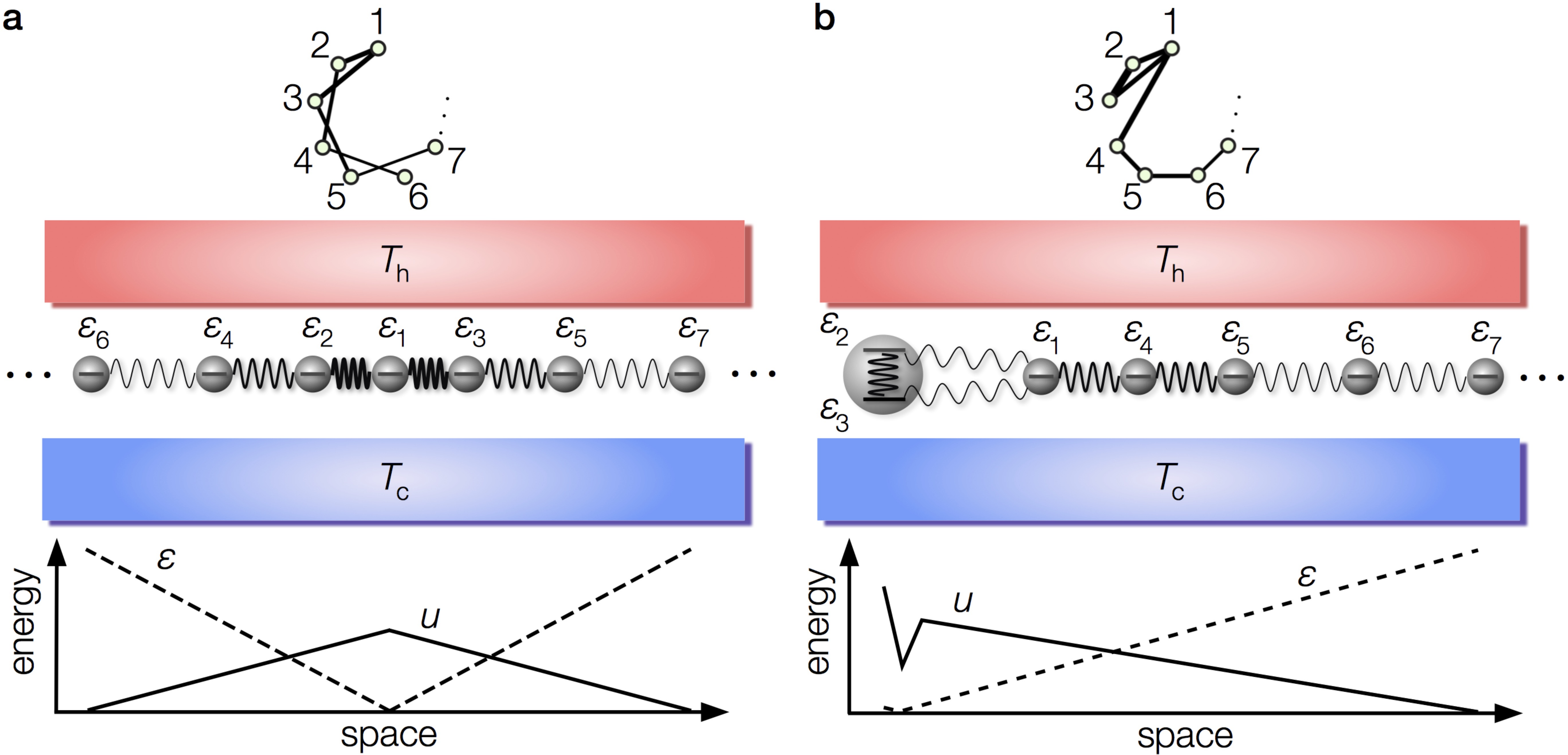}
 \caption{\label{fig5}
Designing the highest-power nanoscale heat engines with quantum-dot arrays.  Illustrations of specific array configurations that can achieve the maximum-power heat engines. For the sake of concreteness, the single-electron energies are assumed to be equal-spacing, i.e., $\epsilon_{l}=l\Delta$. 
 (a) Each quantum dot has a single level and exchanges electrons with reservoirs at temperatures $T_{\rm h,c}$. Quantum dots are linearly aligned from the center in order of increasing energies. Distances are controlled in such a way that nearest-neighbor interactions progressively weaken as distance from the center (see the bottom illustration). The corresponding network topology of the interaction is  shown in the top part of the figure. (b) The left most dot has two levels, while the other dots have a single level and are linearly aligned in order of increasing energies. Distances are controlled in such a way that the interdot (nearest-neighbor) interaction between two leftmost dots is weak while the other interactions progressively weaken as distance from the left (see the bottom illustration). The corresponding network topology of the interaction is also shown in the top part of the figure.
}
\end{figure*}

{\it Discussion.---} 
Our findings can be of experimental relevance to nanoscale thermoelectric described by the classical master equation. As a concrete realization, one may use semiconductor quantum dots embedded into an insulator weakly connected with metallic electrodes. An amorphous insulator (such as SiO$_{2}$) can be used for this purpose since it has low phonon conductivity and its high-potential barrier can block interdot electron hopping \cite{LP00}. Interaction strengths should be controlled by varying the dot distances. 
For typical experimental conditions, the temperature is $1$-$30\,$meV, quantum dots with nanometre size have interaction that is the order of $10$-$100\,$meV, and tunneling rates are $\sim\! 10\,$GHz, which are well within the sequential regime. 

To implement the heat engine supplying the maximum power, we propose two specific examples of quantum-dot arrays with nearest-neighbor interactions as illustrated in Fig.~\ref{fig5}. This type of systems might be engineered by applying manipulation capabilities of nanoporous molecular structures \cite{FK11,IPZ17} or nanoparticles \cite{SKB15}. The configurations in Fig.~\ref{fig5} can attain the conditions (i)-(iii) discussed above  and thus realize the highest-power heat engines (see Supplementary Materials). 
More generally, a guiding principle of realizing the highest-power machine is to make interactions among low (high) single-electron levels strong (weak) to compensate energy differences and make hole excitations degenerate as much as possible. 
While it may in practice be challenging to achieve the ideal limit $N_f\to\infty$, the proposed configurations can still attain high efficiency even at modest system sizes. For instance, assuming $N_f=20$ and the scaling of power $P\propto N_f^{0.6}$, one can achieve the efficiency $\eta/\eta_{\rm C}\simeq 0.90$ with controlling the  chemical potential $\delta\mu$ at the level of $\sim$8\% accuracy of the stopping value $S\delta T$.

It remains as an intriguing open question whether or not further complexities such as nonlinear effects \cite{BS12}, time-reversal symmetry breaking \cite{GB11,BK13} and quantum coherence \cite{MOS11} in nanoscale heat engines allow one to attain better efficiency and power beyond the ones found here. For instance, genuine quantum many-body effects, e.g., the Kondo physics \cite{JK64}, require an explicit account of the entanglement between a system and reservoirs. 
We also remark that the framework developed here for reinforcement learning of network topology on discrete physical states is not restricted to thermoelectric. For example, it can be extended to assembled biomolecules \cite{CJW18}, chemical reactions \cite{DA12} and photovoltaic devices \cite{RDS04} to explore optimal properties of certain objectives such as mobility, reaction yield and rectifications as well as efficiency and power. 

We are grateful to Kensuke Kobayashi, Eiji Saitoh, Junichiro Shiomi, Naoto Shiraishi and Masahito Ueda for valuable comments. Y.A. acknowledges support from the Japan Society for the Promotion of Science (JSPS) through Grant No.~JP16J03613, and Harvard University for hospitality. T.S. acknowledges support from JSPS through Grant Nos.~JP16H02211 and 19H05796.

\widetext
\pagebreak
\begin{center}
\textbf{\large Supplementary Materials}
\end{center}

\renewcommand{\theequation}{S\arabic{equation}}
\renewcommand{\thefigure}{S\arabic{figure}}
\setcounter{equation}{0}
\setcounter{figure}{0}

\subsection{Stochastic thermodynamics of nanothermoelectric heat engines}
Here we describe the master-equation framework to describe the dynamics of nanothermoelectric heat engines. We consider a many-body system governed by an interacting Hamiltonian
\eqn{\label{Hams}
{H}=\sum_{l}(\epsilon_{l}-v_{g})n_{l}+\frac{1}{2}\sum_{l\neq m}w_{lm}n_{l}n_{m},
}
where $\epsilon_l$ is single-particle energy of mode $l=1,2,\ldots,N_f$, $v_g$ is the ground voltage, and $w_{lm}\geq 0$ denotes a two-body interaction between electrons occupying single-particle levels $l$ and $m$. 
Corresponding to different sets of the occupation numbers  $n_l=0,1$ (i.e., Fock states),
 there are $2^{N_f}$ states which we label by $a,b,\ldots$. The system is connected to hot (h) and cold (c) reservoirs. We consider the sequential regime \cite{CWB90}, where the dynamics can be described by the master equation \cite{US12,GB17}
  \eqn{
 \frac{dp_{a}}{dt}=\sum_{b}W_{ab}p_{b},\;\;W_{ab}=\Gamma_{ab}-\delta_{ab}\sum_{d}\Gamma_{db},\;\;\Gamma=\Gamma^{{\rm h}}+\Gamma^{{\rm c}},
  }
where $p_a$ denotes the probability distribution in the energy eigenbasis (i.e., the Fock basis), and $W_{ab}$ is the $2^{N_f}\times 2^{N_f}$ transition matrix. The elements of tunneling matrices $\Gamma^{\rm h,c}$ associated with two reservoirs are given by
\eqn{
\Gamma_{ab}^{i}=\gamma_{i}f(\delta s_{ab}^{i}),\;\;f(x)=\frac{1}{1+e^{x}},\;\;\delta s_{ab}^{i}=\frac{E_{a}-E_{b}}{k_{{\rm B}}T_{i}}+(N_{a}-N_{b})\left(-\frac{\mu_{i}}{k_{{\rm B}}T_{i}}\right),
}
where $f(x)$ is the Fermi distribution, $\delta s_{ab}^i$ describes the entropy production associated with the transition from state $b$ to $a$ via reservoir $i={\rm h,c}$, $E_a$ is a many-body eigenenergy of the Hamiltonian~\eqref{Hams} with respect to state $a$, $N_a$ is the corresponding particle number, and $\mu_i$ and $T_i$ are the chemical potential and temperature of reservoir $i={\rm h,c}$, respectively. We here assume that the tunneling rates are independent of energies, i.e., $\gamma_{\rm h,c}^a=\gamma_{\rm h,c}$ for all $a$. For the sake of simplicity, we do not include transitions between eigenstates with the same particle number; such a transition can be relevant when the electron-phonon interaction becomes important \cite{XZ10,JHJ12}. 
This assumption together with the fact that transport occurs due to single-electron tunnelings to reservoirs can uniquely fix the connectivity of the transition matrix $W$. 
We note that the master equation has a unique steady-state solution $Wp^{{\rm ss}}=0$ since the transfer matrix satisfies the ergodicity \cite{JS76}. 

To calculate the transport coefficients, we focus on the regime $\delta T=T_{{\rm h}}-T_{{\rm c}}\ll T_{{\rm h}}$ and  $|\delta\mu|=|\mu_{{\rm h}}-\mu_{{\rm c}}|\ll k_{{\rm B}}T_{{\rm h}}$, and denote $T=T_{{\rm h}}\simeq T_{{\rm c}}$. We set  $\mu_{{\rm c}}\simeq\mu_{{\rm h}}=0$ without loss of generality. The figure of merit $ZT$ and the power factor $Q$ are then given by the Onsager coefficients as
\eqn{
ZT=\frac{\sigma S^{2}T}{\kappa}=\frac{L_{12}L_{21}}{{\rm det}({\bf L})},\;\;Q=\sigma S^{2}=\frac{L_{12}^{2}}{T^{3}L_{11}},\;\;\left(\begin{array}{c}
J^{\rm h}\\
J_{q}^{\rm h}
\end{array}\right)={\bf L}\left(\begin{array}{c}
\delta\mu/T\\
\delta T/T^{2}
\end{array}\right),\label{onsager}
}
where $\sigma$ is the conductance, $S$ is the Seebeck coefficient, $\kappa$ is the thermal conductance, and ${\bf L}$ is the Onsager matrix. The charge and heat currents into the system from the hot reservoir are denoted by $J^{\rm h}$ and $J_q^{\rm h}$, respectively. The present system can operate as a heat engine when we set $\delta\mu<0$ if $S>0$ while $\delta\mu>0$ if $S<0$.  We numerically obtain the Onsager coefficients (and thus $ZT$ and $Q$ accordingly) from Eq.~\eqref{onsager} by calculating the steady currents $(J^{{\rm ss}},J_{q}^{{\rm ss}})$ for two different steady-state solutions corresponding to $\delta T=0,\delta\mu\neq0$ and $\delta T\neq0,\delta\mu=0$. The nonzero values of $\delta T$ and $\delta \mu$ are kept sufficiently small in such a way that the linear response theory is valid. 

\subsection{Global optimization for the best heat engines} 
We here describe in detail the global optimization algorithm used to identify the best nanoscale heat engines. We start from finding the engine that achieves the maximum possible power factor. This has been done by using the differential evolution \cite{RS95,FN09} 
to optimize the objective function $Q$ with respect to the parameters ${\cal W}=\left\{v_g,\{w_{lm}\}_{l>m}\right\}$. 

Specifically, we first randomly generate a population of $N_p$ $d$-dimensional real-valued vectors $\{{\cal W}_{t=0}^i\}$ with $i=1,2,\ldots,N_p$. Here, $d=1+N_{f}(N_{f}-1)/2$ is the number of variables in the present problem. 
At each iteration step $t$, we create a mutant vector ${\cal V}_{t}^{i}\in\mathbb{R}^{d}$ with $i=1,2,\ldots,N_p$ by 
\eqn{
{\cal V}_{t}^{i}={\cal W}_{t}^{k}+F({\cal W}_{t}^{l}-{\cal W}_{t}^{m}),
} 
where $k,l,m\neq i$ are mutually exclusive integers randomly chosen from $[1,N_{p}]$, ${\cal W}_{t}^{k}$ is a vector sampled from the population at step $t$, and $F>0$ is the scaling factor.
We next prepare an offspring vector ${\cal X}_{t}^{i}\in\mathbb{R}^{d}$ by randomly choosing an integer $k$ from $[1,d]$ and then applying the following rule to each element of ${\cal X}_{t}^{i}$:
\eqn{
({\cal X}_{t}^{i})_{j}=\begin{cases}
\begin{array}{cc}
({\cal V}_{t}^{i})_{j} & {\rm if\;}j=k\;{\rm or}\;{\rm rand}_{i,j}[0,1]\leq C_{r}\\
({\cal W}_{t}^{i})_{j} & {\rm otherwise}
\end{array}\end{cases},
}
where rand$_{i,j}$[0,1] denotes a randomly generated number from [0,1] for each pair of $(i,j)$ and $C_{r}>0$ is the crossover factor. 
We then update all the vectors $\{{\cal W}_{t}^i\}$ in the population according to the rule
\eqn{
{\cal W}_{t+1}^{i}=\begin{cases}
\begin{array}{cc}
{\cal X}_{t}^{i} & {\rm if\;}Q({{\cal X}_{t}^{i}})\geq Q({{\cal W}_{t}^{i}})\\
{\cal W}_{t}^{i} & {\rm otherwise}
\end{array}\end{cases}.
}
To improve the convergence, we restrict the search space by setting the lower and upper bounds on the parameters as $0\leq w_{lm}\leq w_{\rm max}$ and $-v_{\rm max}\leq v_{g}\leq v_{\rm max}$; we set $w_{\rm max}=v_{\rm max}=50k_{{\rm B}}T$ for all the results provided in the main text. The population number $N_p$ is set to be $N_p=2d$. 
We find it useful to implement the heuristics proposed in Ref.~\cite{YW11}, 
where the scaling factor and the crossover factor $(F,C_{r})$ are chosen uniformly at random from three choices $(1,0.1)$, $(1,0.9)$ and $(0.8,0.2)$ at each iteration. The choice of $(1,0.9)$ leads to large perturbations on the donor vectors and thus expedites the exploration of the search space while the other two choices expedite the exploitation of the search space. 
In this way, we can find the optimal parameters ${\cal W}_{Q}^*$ for the highest-power heat engine, which provides an unambiguous element of the Pareto-optimal solutions \cite{YS85}. 
It thus allows us to determine the Pareto front on the $Q$-$ZT$ plane by starting the search of the front with setting ${\cal W}_{Q}^*$ to be the initial point. The algorithm is based on the alternate search of the feasible parameter region and the higher values of the objective functions $Q$ and $ZT$   \cite{CAL11}. Finally, to obtain the Pareto front on the power-efficiency plane, we generate a large number of loops according to the linear-response formula \cite{GB17}
\eqn{
\frac{\eta(P)}{\eta_{{\rm C}}}=\frac{P/(Q\delta T^{2}/4)}{2\left[1+2/ZT\mp\sqrt{1-P/(Q\delta T^{2}/4)}\right]}
}
with substituting the values of $(Q^*,ZT^*)$ for the Pareto-optimal solutions into it. The envelope curve of these loops provide the Pareto front on the $P$-$\eta$ plane (see, for example, Fig.~2(a) in the main text). 

Full training for the interacting case with the largest system size has been done by running the algorithm within about 3 days on a single 24-core CPU machine. 
Most of the time is spent on the diagonalization of the transition matrix to evaluate the fitness values of  machines in a large population at each iteration. Thus, the exponential computational cost for the matrix diagonalization currently limits the available number of single-particle levels $N_f$. A reduction of computational times could be made possible by applying the efficient solvers of many-body problems \cite{GV04,YA18} or by parallelizing the code over large-core CPU+GPU machines.

\subsection{Nonconvex landscape and the failure of the local algorithms} 
\begin{figure}[t]
\includegraphics[width=160mm]{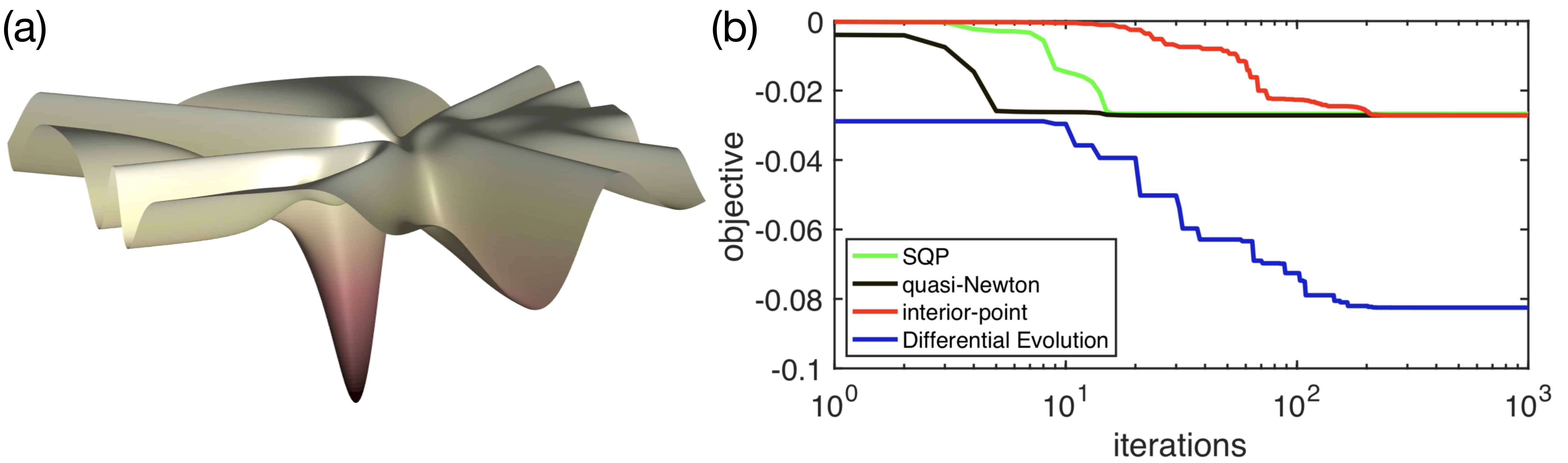}
\caption{\label{figs1}
Visualization of the nonconvex landscape and the failure of the greedy algorithms.  (a) The optimization surface of (minus) the power factor $-Q$ around the global-optimum solution ${\cal W}_{Q}^*$ positioning at the center. We here project the high-dimensional optimization landscape onto the two-dimensional surface based on the approach used in Ref.~\cite{IJG15}. (b) The convergence behaviour of the optimization of the objective function $-Q$ for the sequential quadratic programming (SQP) (green),  the quasi-Newton method (black), the interior-point algorithm (red), and the differential evolution (blue).  The parameters are $N_f=4$, $\epsilon_l=l\Delta$ with $\Delta/k_{\rm B}T=3$, $\gamma_{\rm h}=\gamma_{\rm c}\equiv \gamma$, and we plot the power factor in the unit of $k_{\rm B}\gamma/T$.
}
\end{figure}
\noindent
When the optimization landscape is smooth and local traps therein are negligible, 
the greedy algorithms should be the primary choice since they can usually reach to the global optimum with faster convergence rate than that for the global (derivative-free) algorithms.
We here briefly mention the difficulties of applying such greedy (gradient-based) optimization algorithms to identify the best interacting nanoscale heat engines. To demonstrate this, we show typical training processes in optimizing the minus of the power factor, $-Q$, in Supplementary  Figure~\ref{figs1}(b). The result is presented for a small system size with $2^{N_f}=16$ states. We here apply three standard local optimization algorithms: the sequential quadratic programming (SQP) (green) \cite{HW83},  the quasi-Newton method (black) \cite{FR63}, and the interior-point algorithm (red) \cite{WRA06}.  As inferred from the figure, all of three greedy algorithms overwhelmingly fail to identify the global-optimum solution and are easily trapped by the local optima. For instance, through the training processes shown, the value of the suboptimal power factor reached by the greedy algorithms is almost the same as in that of the best power factor achieved by the noninteracting engine (see e.g., the left bottom panel in Fig.~2 in the main text). This observation can be understood from the fact that the suboptimal engines are able to activate only a very few number (one or two) of transfer edges in the state-transfer network (cf. panels (a) and (d) in Fig.~2 in the main text), resulting in low (yet locally optimal) values of the power factor.   

Numerical evidence of the failure of the greedy algorithms indicates the presence of a large number of proliferated local optima in the nonconvex landscape of the search space. To demonstrate this more explicitly, we visualize the optimization landscape based on the approach previously applied to the problem of optimizing the deep neural network \cite{IJG15}. 
Specifically, we randomly choose vectors $\phi,\psi\in {\mathbb R}^d$ with $d=1+N_{f}(N_{f}-1)/2$ and project the landscape onto the two-dimensional surface via introducing the objective
\eqn{
L(\alpha,\beta)=Q({\cal W}_{Q}^{*}+\alpha\phi+\beta\psi),
}
where ${\cal W}_{Q}^*$ is the global-optimum solution and $\alpha,\beta$ are real parameters. Supplementary Figure~\ref{figs1}(a) shows a typical example of the obtained two-dimensional projected optimization landscape. 
It exhibits the dramatic nonconvexities and, in many regions, the surface gradients do not point toward the global optimum positioned at the center. Thus, most trials of the greedy algorithms to find the global optimum converge to local traps in the landscape. 
It is this complex nature of the optimization landscape that leads to the failure of the greedy (gradient-based) algorithms as demonstrated in Supplementary Figure~\ref{figs1}(b). In contrast, the differential evolution (detailed in the previous section) is known as one of the most powerful approaches to find the global solution in the high-dimensional nonconvex landscape \cite{FN09}, and has indeed been able to find the globally optimal power factor (see the blue line in Supplementary Figure~\ref{figs1}(b)).

We finally mention the difficulty of applying the brute-force search to the present problem. Since the brute-force approach requires the search over all the possible patterns of the interaction network, one has to discretize the range of each continuous parameter in ${\cal W}$ with a total amount $L_{\rm dis}$ of bins. Thus, the resulting number of the required trials becomes exponentially large, i.e., it scales as $\propto (L_{\rm dis})^{d}$ with $d=N_f(N_f-1)/2+1$. Since  the computational cost for each trial also grows exponentially with $N_f$ due to the need of diagonalizing the transition matrix $W$, the brute-force search leads to the double-exponential growth of the numerical cost as increasing the system size. We note that the resulting solutions may still not reach the global optimum since the search is (by nature) nonexhaustive due to the discretization of the continuous parameters. 

\subsection{Details of the parameters for the asymptotic Carnot engines at nonzero power} 
We here mention the concrete examples of the interaction parameters appropriate for realizing the highest-power heat engines. They can asymptotically achieve the Carnot efficiency at nonzero and stable power in the thermodynamic limit $N_f\to \infty$ owing to the diverging power factor $Q$.
Firstly, as the parameters appropriate for the one-dimensional chain configuration with single-particle energies $\epsilon_l=l\Delta $ (cf. Fig.~5(a) in the main text), one may use
\eqn{
w_{12}&=&2\left[\frac{N_{f}+1}{4}\right]\Delta,\\
w_{l,l+2}&=&\left(\left\lceil\frac{N_{f}-1}{2}\right\rceil-\left[\frac{l}{2}\right]\right)\Delta\;\;\;{\rm for}\;\;\;l=1,2,5,6,\ldots,\\
w_{l,l+2}&=&w_{12}-2\left\lceil\frac{l}{4}\right\rceil\Delta\;\;\;{\rm for}\;\;\;l=3,4,7,8,\ldots,
}
and set the other interaction parameters to be zero. Here, $[\cdot]$ is the Gauss symbol and $\lceil\cdot\rceil$ is the ceiling function. 
Secondly, as the parameters appropriate for the semi-infinite chain configuration (cf. Fig.~5(b) in the main text), one may use
\eqn{
w_{23}&=&\frac{\lceil3N_{f}/2\rceil-6}{2}\Delta,\;\;w_{12}-\Delta=w_{13}=\frac{[N_{f}/2]\Delta}{2},\;\;w_{14}=(\lceil N_{f}/2\rceil-2)\Delta,\\
w_{l,l+1}&=&\left[\frac{N_{f}-l}{2}\right]\Delta\;\;\;{\rm for}\;\;\;l=4,5,\ldots. 
}
One can check that these two sets of the parameters satisfy the optimality conditions for the maximum-power engines (proposed in the main text), which include (i) single-hole excitation energies are degenerate and (ii) the number of nonzero parameters in  $\{w_{lm}\}_{l>m}$ is minimal.

\end{document}